\documentclass[letterpaper,english,notitlepage,twocolumn, superscriptaddress]{revtex4-1}
\usepackage[T1]{fontenc}
\usepackage[latin9]{inputenc}
\usepackage{amsmath}
\usepackage{amssymb}
\usepackage{graphicx}

\makeatletter

\pdfpageheight\paperheight
\pdfpagewidth\paperwidth

\usepackage[colorlinks,citecolor=blue,linkcolor=blue]{hyperref}

\makeatother

\usepackage{babel}
\begin{document}
\title{Heat statistics in the relaxation process of the Edwards-Wilkinson
elastic manifold}
\author{Yu-Xin Wu}
\affiliation{School of Physics, Peking University, Beijing, 100871, China}
\author{Jin-Fu Chen}
\affiliation{School of Physics, Peking University, Beijing, 100871, China}
\author{Ji-Hui Pei}
\affiliation{School of Physics, Peking University, Beijing, 100871, China}
\author{Fan Zhang}
\affiliation{School of Physics, Peking University, Beijing, 100871, China}
\author{H. T. Quan}
\thanks{Corresponding author: htquan@pku.edu.cn}
\affiliation{School of Physics, Peking University, Beijing, 100871, China}
\affiliation{Collaborative Innovation Center of Quantum Matter, Beijing, 100871,
China}
\affiliation{Frontiers Science Center for Nano-optoelectronics, Peking University,
Beijing, 100871, China}
\date{\today}
\begin{abstract}
The stochastic thermodynamics of systems with a few degrees of freedom
has been studied extensively so far. We would like to extend the study
to systems with more degrees of freedom and even further\textendash continuous
fields with infinite degrees of freedom. The simplest case for a continuous
stochastic field is the Edwards-Wilkinson elastic manifold. It is
an exactly solvable model of which the heat statistics in the relaxation
process can be calculated analytically. The cumulants require a cutoff
spacing to avoid ultra-violet divergence. The scaling behavior of
the heat cumulants with time and the system size as well as the large
deviation rate function of the heat statistics in the large size limit
is obtained.
\end{abstract}
\maketitle

\section{Introduction}

Historically, people studied thermodynamics in macroscopic systems
such as ideal gas with up to $10^{23}$ molecules. Due to the huge
number of degrees of freedom in the macroscopic scale, it is impossible
to extract the trajectories of individual particles explicitly. Hence
it is not possible to study thermodynamics of macroscopic systems
in arbitrary far from equilibrium processes. Nevertheless, for mesoscopic
systems with only a few degrees of freedom, stochastic dynamics (Langevin
equation, Fokker-Planck equation, master equation) provides detailed
information about the system. Prominent examples of mesoscopic systems
include colloidal particles, macromolecules, nanodevices and so on
\citep{Livi_book_2017,Peliti_book_2021}. In all these examples, researchers
focus on the dynamics of a few degrees of freedom of the system while
coarse-graining all the degrees of freedom of the reservoir. Mesoscopic
systems can be driven out of equilibrium by external driving, for
instance, by varying the temperature or by controlling them with optical
tweezers \citep{Hummer_PNAS_2001,Liphardt_Sci_2002,Wang_PhysRevLett_2002,Blickle_PhysRevLett_2006,Douarche_PhysRevLett_2006,Harris_PhysRevLett_2007,Imparato_PhysRevE_2007,Toyabe_NatPhys_2010,Gupta_NatPhys_2011,Alemany_NatPhys_2012,Gieseler_NatNanoTech_2014,Jun_PhysRevLett_2014,Koski_PhysRevLett_2014,Lee_PhysRevLett_2015,Martinez_NatPhys_2015,Hoang_PhysRevLett_2018}. 

With the equation of motion, e.g., Langevin equation, Fokker-Planck
equation or master equation, researchers are able to establish a framework
of thermodynamics for mesoscopic systems in arbitrary far from equilibrium
processes. This is stochastic thermodynamics in which thermodynamic
quantities such as work, heat and entropy production in nonequilibrium
processes have been explored extensively in both classical and quantum
realms \citep{Jarzynski_PhysRevLett_1997,Mazonka_Arxiv_1999,Narayan_JPhysA_2003,Speck_PhysRevE_2004,Zon_PhysRevE_2004,Lua_JPhysChemB_2005,Speck_EuroPhysJB_2005,Taniguchi_JStatPhys_2006,Imparato_PhysRevE_2007,Quan_PhysRevE_2008,Engel_PhysRevE_2009,Fogedby_JPhysA_2009,Minh_PhysRevE_2009,Chatterjee_PhysRevE_2010,GomezSolano_PhysRevLett_2011,Nickelsen_EuroPhysJB_2011,Speck_JPhysA_2011,Kwon_PhysRevE_2013,JimenezAquino_PhysRevE_2013,Ryabov_JPhysA_2013,Jarzynski_PhysRevX_2015,Salazar_JPhysA_2016,Zhu_PhysRevE_2016,Funo_PhysRevLett_2018,Funo_PhysRevE_2018,Hoang_PhysRevLett_2018,Pagare_PhysRevE_2019,Fogedby_JStatMech_2020,Chen_Entropy_2021,Gupta_PhysRevE_2021,Chen_PhysRevE_2023,Paraguassu_PhysicaA_2023}.
In the study of work or heat distribution for extreme nonequilibrium
processes, rare events with exponentially small probabilities have
dominant contributions making finite sampling error particularly serious.
Hence previous studies, be it experimental or computer simulations,
are predominantly for small systems, i.e., those with a few degrees
of freedom \citep{Hartmann_PhysRevE_2014}. Nevertheless, systems
with a few degrees of freedom are too special. Therefore it is desirable
to extend the study of stochastic thermodynamics to more complicated
systems. We thus would like to extend the studies to systems with
more degrees of freedom, for example, stochastic fields. Hopefully
in some exactly solvable model we can obtain analytical results about
work and heat distribution. These rigorous results about work or heat
distribution in systems with many degrees of freedom not only have
pedagogical value but also may bring some insights to the understanding
of thermodynamics in extreme nonequilibrium processes, as P. W. Anderson
once advocated, ``More is different'' \citep{Anderson_Sci_1972}.
While many researchers are interested in the dynamic properties of
stochastic fields \citep{Forrest_PhysRevLett_1990,Antal_PhysRevE_1996,Racz_SPIE_2003,Vvedensky_PhysRevE_2003,Bustingorry_JStatMech_2007},
less research is carried out from the perspective of stochastic thermodynamics
except \citep{Mallick_JPhysA_2011,Wio_FrontPhys_2017,Rodriguez_PhysRevE_2019,Wio_Chaos_2020,Wio_JStatMech_2020}
so far as we know.

In this article we study the thermodynamics of an elastic manifold
whose underlying dynamics is described by the Edwards-Wilkinson (EW)
equation \citep{Edward_Wilkinson_ProcRoySocLon_1982}

\begin{equation}
\partial_{t}h(\boldsymbol{x},t)=\nu\nabla^{2}h(\boldsymbol{x},t)+\xi(\boldsymbol{x},t).\label{eq:EW}
\end{equation}
where $h(\boldsymbol{x},t)$ is the local height at spatial point
$\boldsymbol{x}$ at time $t$, $\nu$ is the diffusive coefficient
and $\xi(\boldsymbol{x},t)$ is the Gaussian white noise. 

The problem we analyze is the relaxation of an elastic manifold described
by the EW equation. The elastic manifold is initially put in contact
with a heat reservoir at the inverse temperature $\beta'$. After
initial equilibration with the first heat reservoir at $\beta'$ the
system is detached from it, and is put in contact with a second heat
reservoir at the inverse temperature $\beta$. The manifold subsequently
tries to adapt to the working temperature \citep{Bustingorry_JStatMech_2007}.
The relaxation is characterized by the stochastic heat absorbed from/released
into the surrounding reservoir during a period of time $\tau$. We
are interested in the average and fluctuation of the heat in such
a process. We find several generic properties of the average and fluctuating
heat in the relaxation process of the EW elastic manifold. By employing
the Feynman-Kac method \citep{Chen_Entropy_2021,Limmer_EurPhysJB_2021},
we obtain analytical results of the characteristic function of heat
for the EW model during an arbitrary relaxation period $\tau$ with
an arbitrary diffusive coefficient $\nu$ and analyze the scaling
behavior of the cumulants of heat with time. Analytical results of
the heat statistics bring important insights into understanding the
fluctuating property of heat in such a concrete and exactly solvable
model. We also verify from the analytical results that the heat statistics
satisfy the fluctuation theorem of heat exchange \citep{Jarzynski_PhysRevLett_2004}.
The large deviation rate function of heat statistics in the large
size limit is also analyzed.

The rest of this article is organized as follows. In Section \ref{sec:The-model}
we introduce the EW model. In Section \ref{sec:Heat-Statistics} we
define the stochastic heat and obtain analytical results of the characteristic
function of heat using the Feynman-Kac approach. We also compute the
cumulants of heat and discuss their scaling behavior with time and
the system size. Conclusions are given in Section \ref{sec:Conclusion}.

\section{The model\label{sec:The-model}}

A $d$-dimensional elastic manifold, with finite size $2L$ in each
direction, joggles under thermal noise. Its local height $h(\boldsymbol{x},t)$
at spatial point $\boldsymbol{x}$ at time $t$ evolves according
to the EW equation Eq. (\ref{eq:EW}) which takes the form of a multivariable
overdamped Langevin equation \citep{Livi_book_2017}. The thermal
noise $\xi(\boldsymbol{x},t)$ is white in nature, i.e., $\langle\xi(\boldsymbol{x},t)\rangle=0,$
$\langle\xi(\boldsymbol{x},t)\xi(\boldsymbol{x}',t')\rangle=\Gamma\delta(\boldsymbol{x}-\boldsymbol{x}')\delta(t-t'),$
with amplitude \textbf{$\Gamma=2/\beta$}. The EW energy is just that
of a massless field with Hamiltonian $H_{S}=\nu\int d\boldsymbol{x}(\nabla h(\boldsymbol{x},t))^{2}/2$.
Here the subscript $S$ refers to the system.

Initially, the system is prepared in an equilibrium state with the
inverse temperature $\beta'$ characterized by a Gibbs-Boltzmann distribution
in the configuration space, i.e., the probability ${\cal P}(h,t)$
to find the system in the configuration $\{h(\boldsymbol{x},t)\}$
is the Gibbs-Boltzmann distribution

\begin{equation}
{\cal P}(h,0)={\cal N}'^{-1}\exp\Big[-\beta'\cdot\frac{\nu}{2}\int d\boldsymbol{x}\Big(\nabla h(\boldsymbol{x},0)\Big)^{2}\Big]
\end{equation}
where ${\cal N}'$ is the normalization constant 

\begin{align}
{\cal N}' & =\int dh(\boldsymbol{x},0)\exp\Big[-\beta'\cdot\frac{\nu}{2}\int d\boldsymbol{x}\Big(\nabla h(\boldsymbol{x},0)\Big)^{2}\Big].\label{eq:normalization_constant}
\end{align}
Here the integration in the normalization constant is taken over all
possible initial configurations while the one in the exponential factor
is taken over all spatial points. 

After initial equilibration, the system is detached from the first
heat reservoir, and is placed in contact with a second heat reservoir
at the inverse temperature $\beta$, which is different from $\beta'$.
The elastic manifold subsequently relaxes towards the equilibrium
state at temperature $\beta$ since no external driving is involved.
The heat absorbed/released is a fluctuating variable for the system
undergoing stochastic motion. We are interested in the heat statistics
in such a relaxation process. 

For a finite-size manifold we take periodic boundary conditions along
each $\boldsymbol{x}$ direction. Following Refs. \citep{Antal_PhysRevE_1996,Livi_book_2017}
we employ a Fourier representation of the height field

\begin{equation}
h(\boldsymbol{x},t)=\frac{1}{(2\pi)^{d}}\underset{\boldsymbol{q}}{\sum}e^{i\boldsymbol{q}\cdot\boldsymbol{x}}h_{\boldsymbol{q}}(t),
\end{equation}

\begin{equation}
h_{\boldsymbol{q}}(t)=\int d\boldsymbol{x}e^{-i\boldsymbol{q}\cdot\boldsymbol{x}}h(\boldsymbol{x},t),
\end{equation}
where $\boldsymbol{q}$ represents a wavevector with $q_{j}=n_{j}\pi/L\ (j=x,y,z\dots,\ n_{j}=\pm1,\pm2...$
and $h_{\boldsymbol{q}=\boldsymbol{0}}(t)=0$ for all time $t)$ \citep{Bustingorry_JStatMech_2007}.\\
The evolution of the Fourier component is given by

\begin{equation}
\frac{\partial h_{\boldsymbol{q}}(t)}{\partial t}=-\nu q^{2}h_{\boldsymbol{q}}(t)+\xi_{\boldsymbol{q}}(t),
\end{equation}

\begin{equation}
\langle\xi_{\boldsymbol{q}}(t)\rangle=0,
\end{equation}

\begin{align}
\langle\xi_{\boldsymbol{q}}(t)\xi_{\boldsymbol{q'}}(t')\rangle & =\frac{2}{\beta}(2\pi)^{d}\delta(t-t')\delta_{\boldsymbol{q},-\boldsymbol{q}'}.
\end{align}
The normalization constant in Eq. (\ref{eq:normalization_constant})
can be computed as

\begin{align}
{\cal N}' & =\int d\{h_{\boldsymbol{q}}(0)\}\exp\Big[-\beta'\nu\frac{1}{(2\pi)^{2d}}\underset{\boldsymbol{q}(q_{j}>0)}{\sum}q^{2}h_{\boldsymbol{q}}(0)h_{-\boldsymbol{q}}(0)\Big]\nonumber \\
 & =\underset{\boldsymbol{q}(q_{j}>0)}{\prod}\frac{\pi(2\pi)^{2d}}{\beta'\nu q^{2}}.
\end{align}
where $q^{2}$ stands for the modulus square of $\boldsymbol{q}.$

The probability density of the system state ${\cal P}(h,t)$ evolves
under the governing of the Fokker-Planck equation

\begin{align}
\frac{\partial{\cal P}(h,t)}{\partial t} & =-\int d\boldsymbol{x}\frac{\delta}{\delta h}\Big[\nu\nabla^{2}h(\boldsymbol{x},t){\cal P}(h,t)\Big]\nonumber \\
 & \quad+\frac{\Gamma}{2}\int d\boldsymbol{x}\frac{\delta^{2}}{\delta h^{2}}{\cal P}(h,t).\label{eq:Fokker Planck eq.}
\end{align}
In the Fourier space, the probability of the height field configuration
is the product of the real and the imaginary part over all modes

\begin{equation}
{\cal P}(\{h_{\boldsymbol{q}}\},t)=\underset{\boldsymbol{q}}{\prod}{\cal P}(h_{\boldsymbol{q}},t)=\underset{\boldsymbol{q}}{\sum}{\cal P}(h_{\boldsymbol{q}}^{R},t){\cal P}(h_{\boldsymbol{q}}^{I},t)
\end{equation}
where
\begin{equation}
h_{\boldsymbol{q}}^{R}=\mathrm{Re}(h_{\boldsymbol{q}}),\quad h_{\boldsymbol{q}}^{I}=\mathrm{Im}(h_{\boldsymbol{q}}).
\end{equation}
The Fokker-Planck equation in the Fourier space can be then written
into two independent parts: the real part and the imaginary part \citep{Bettencourt_PhysRevD_2001}

\begin{align}
\frac{\partial{\cal P}(h_{\boldsymbol{q}}^{R,I},t)}{\partial t} & =\frac{(2\pi)^{d}}{2\beta}\frac{\partial^{2}{\cal P}}{\partial(h_{\boldsymbol{q}}^{R,I})^{2}}+\nu q^{2}{\cal P}+\nu q^{2}h_{\boldsymbol{q}}^{R,I}\frac{\partial{\cal P}}{\partial h_{\boldsymbol{q}}^{R,I}}.\label{eq:Fokker-Planck Fourier}
\end{align}

Having introduced the model, in the following we will calculate the
heat statistics in the relaxation process.

\section{Heat Statistics\label{sec:Heat-Statistics}}

In this section we study heat statistics of the EW elastic manifold
in the relaxation process. First, we obtain the analytical results
of heat statistics and verify the fluctuation theorem of heat exchange.
Second, we study the asymptotic behavior of the cumulants. Third,
we calculate the large deviation function of heat statistics in the
large size limit.

\subsection{Characteristic function }

Since no external driving is applied to the system, no work is performed
during the relaxation process. The fluctuating heat $Q$ absorbed
from the heat reservoir equals the energy difference between the initial
and the final states over a time period $\tau$

\begin{equation}
Q=H_{S}(h(x,\tau))-H_{S}(h(x,0)).
\end{equation}
The characteristic function of heat $\chi_{\tau}(u)$ is defined as
the Fourier transform of the heat distribution

\begin{equation}
\chi_{\tau}(u)=\int dQ\exp(iuQ){\cal P}(Q,\tau).
\end{equation}
Here ${\cal P}(Q,\tau)$ stands for the probability of the heat $Q$
transferred from the heat reservoir to the system during the period
of time $\tau$. The characteristic function of heat $\chi_{\tau}(u)$
can be calculated using the Feynman-Kac approach \citep{Chen_Entropy_2021,Limmer_EurPhysJB_2021,Chen_PhysRevE_2023}

\begin{align}
\chi_{\tau}(u) & =\langle\exp(iuQ)\rangle\nonumber \\
 & =\int dhe^{iuH_{S}(h(x,\tau))}\eta(h,\tau)\label{eq:GF_heat_general}
\end{align}
where the probability-density-like function $\eta(h,\tau)$ satisfies
Eq. (\ref{eq:Fokker Planck eq.}) and Eq. (\ref{eq:Fokker-Planck Fourier})
with the initial condition 

\begin{align}
\eta(h,0) & =e^{-iuH_{S}(h(x,0))}{\cal P}(h,0).
\end{align}
The probability-density-like function $\eta(h,\tau)$ is solved in
the Fourier space (See Appendix \ref{sec:Derivation} for detailed
derivation) and we obtain the characteristic function of heat for
the relaxation process over a time period of $\tau$ \begin{widetext}

\begin{equation}
\chi_{\tau}(u)=\beta\beta'\underset{\boldsymbol{q}(q_{j}\geq\frac{\pi}{L})}{\prod}\frac{\exp(2\nu q^{2}\tau)}{-u(i\beta'-i\beta-u)\Big[\exp(2\nu q^{2}\tau)-1\Big]+\beta\beta'\exp(2\nu q^{2}\tau)}.\label{eq:GF_of_heat_result}
\end{equation}
\end{widetext}The wavevector component in each direction only takes
positive discrete values $q_{j}=n_{j}\pi/L,n_{j}=1,2...$

We do the self-consistent check of the analytic result Eq. (\ref{eq:GF_of_heat_result})
from three aspects:\\
1. The distribution of heat satisfies the conservation of probability
\begin{equation}
\chi_{\tau}(0)=1.
\end{equation}
2. One can see the characteristic function of heat exhibits the following
symmetry:
\begin{equation}
\chi_{\tau}(u)=\chi_{\tau}(i\beta'-i\beta-u),
\end{equation}
indicating that the heat distribution satisfies the fluctuation theorem
of heat exchange \citep{Jarzynski_PhysRevLett_2004,Zon_PhysRevE_2004,Chen_PhysRevE_2023}
\begin{align}
\langle e^{iuQ}\rangle & =\langle e^{(-iu+\beta-\beta')Q}\rangle.\label{eq:FT of heat exchange}
\end{align}
By setting $u=0$, we obtain the relation $\chi_{\tau}(i\beta'-i\beta)=1$,
which is exactly the fluctuation theorem of heat exchange in the integral
form $\langle\exp[-(\beta'-\beta)Q]\rangle=1$\citep{Jarzynski_PhysRevLett_2004}.\\
3. In the long time limit $\tau\rightarrow\infty$, the characteristic
function becomes
\begin{align*}
\underset{\tau\to\infty}{\lim}\chi_{\tau}(u) & =\underset{\boldsymbol{q}(q_{j}\geq\frac{\pi}{L})}{\prod}\frac{\beta\beta'}{(u+i\beta)(u-i\beta')}.
\end{align*}
This result, independent of the relaxation dynamics, can be written
in the form
\begin{align}
\underset{\tau\to\infty}{\lim}\chi_{\tau}(u) & =\Big\langle e^{iuH_{S}(h(x,\tau))}\Big\rangle_{\beta}\Big\langle e^{-iuH_{S}(h(x,0))}\Big\rangle_{\beta'}
\end{align}
where the initial distribution (thermal equilibrium with the inverse
temperature $\beta'$) and the final distribution (thermal equilibrium
with the inverse temperature $\beta$) are sampled independently,
reflecting the complete thermalization of the system \citep{Fogedby_JPhysA_2009}.
This result agrees with our intuition.

\begin{figure}
\begin{raggedright}
(a)\\
\par\end{raggedright}
\includegraphics[width=8.5cm]{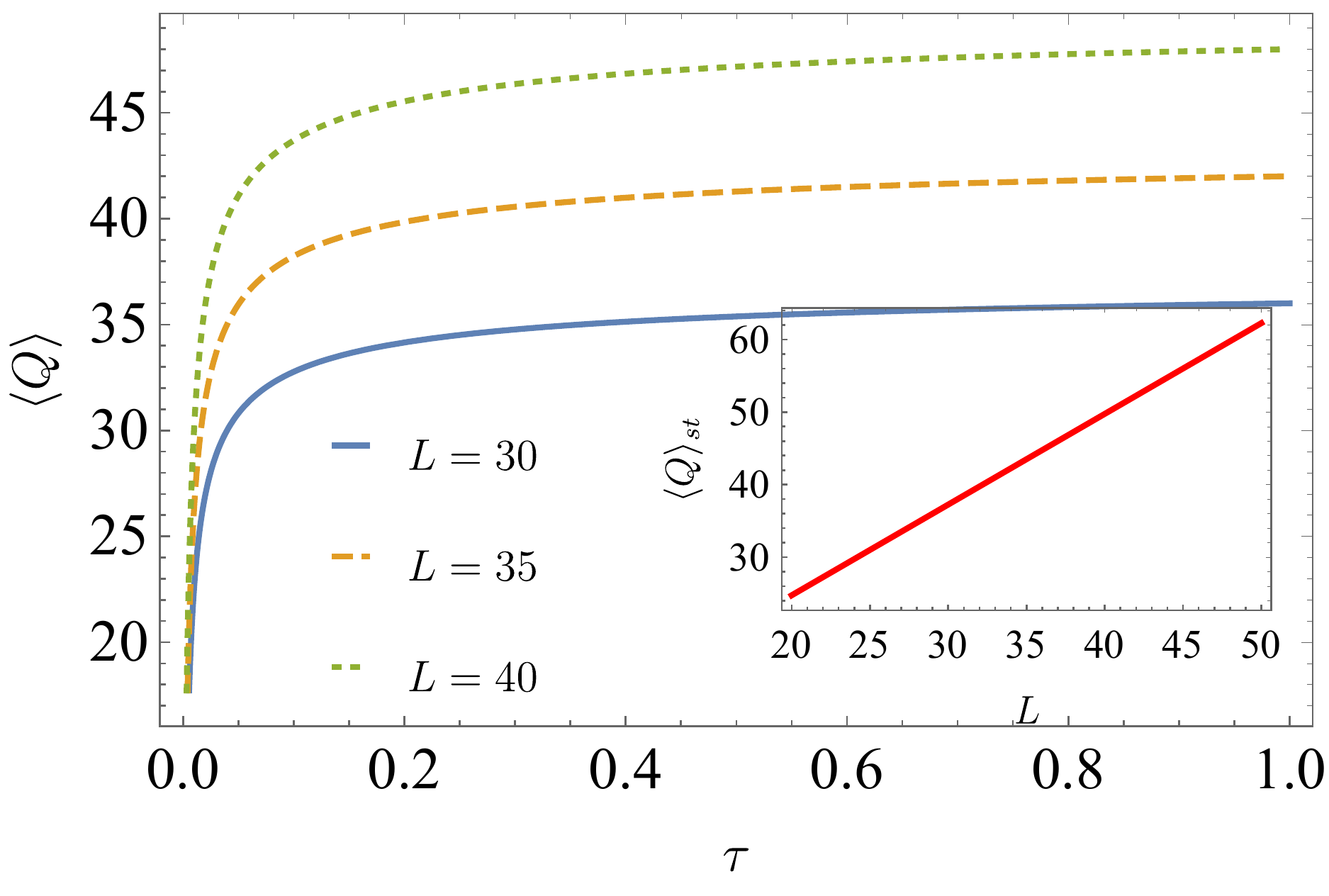}
\begin{raggedright}
(b)\\
\par\end{raggedright}
\includegraphics[width=8.5cm]{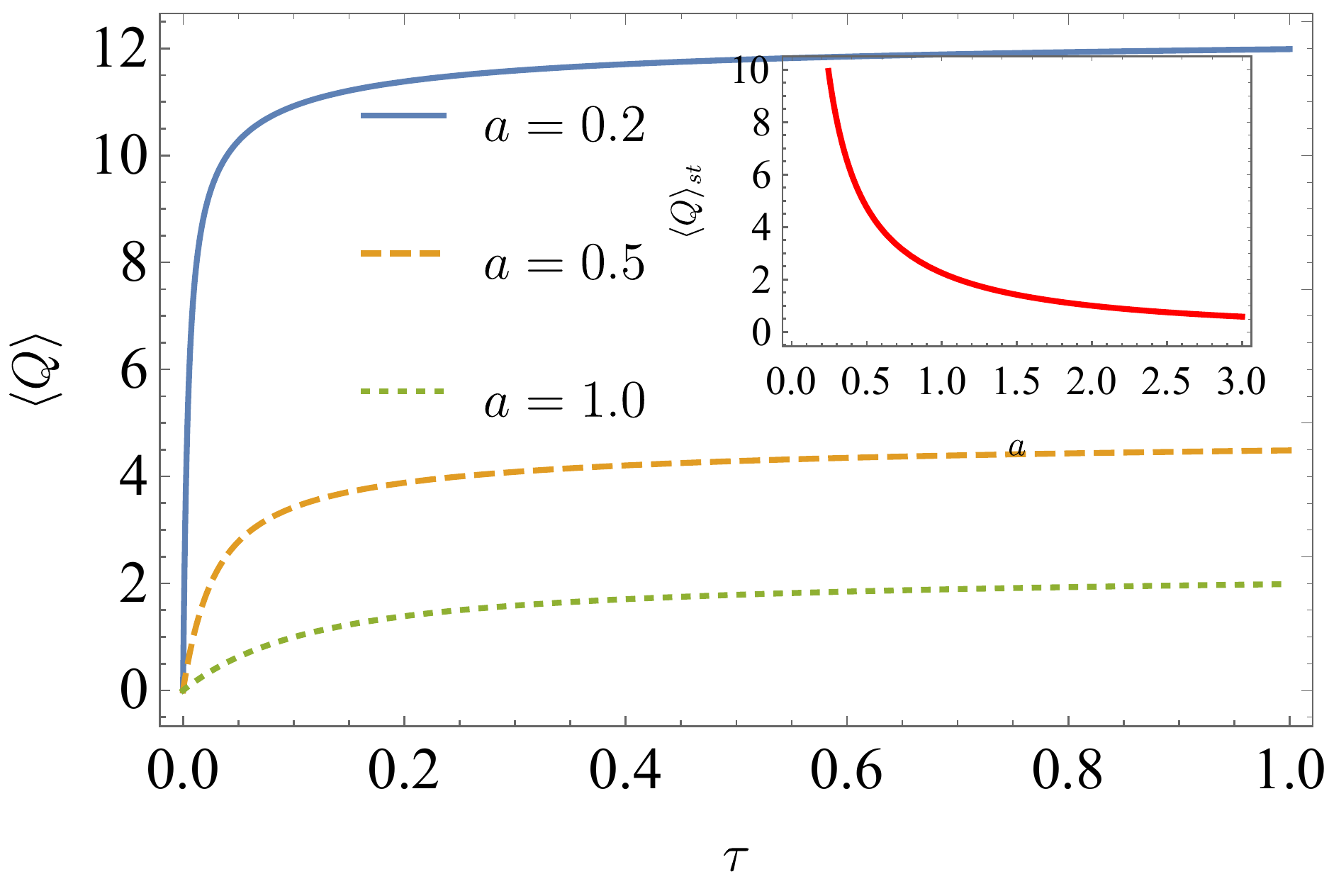}

\caption{{\footnotesize{}Average heat as a function of time. Parameters for
both panels: $d=1,\ \nu=1,\ \beta'=4,\ \beta=2.$ (a) $\langle Q\rangle$
as a function of $\tau$ for three system sizes $L=30,35,40$, fixing
$a=0.2.$ Inset: the saturation value of average heat $\langle Q\rangle_{st}$
as a function of system size $L$. (b) $\langle Q\rangle$ as a function
of $\tau$ for three cutoff spacings $a=0.2,0.5,1.0$, fixing $L=10.$
Inset: the saturation value of average heat $\langle Q\rangle_{st}$
as a function of cutoff spacing $a$. \label{fig:mean_Q}}}
\end{figure}

\subsection{Cumulants }

The cumulants of heat can be derived by taking derivatives of the
logarithm of the characteristic function $\chi_{\tau}(u)$ with respect
to $u$ at $u=0$, with the first cumulant representing the average
heat and the second one standing for the variance.

The average heat is 
\begin{align}
\langle Q\rangle & =\frac{1}{i}\frac{d\ln\chi_{\tau}(u)}{du}|_{u=0}\nonumber \\
 & =\underset{\boldsymbol{q}(\frac{\pi}{a}\geq q_{j}\geq\frac{\pi}{L})}{\sum}\frac{\Big[1-\exp(-2\nu q^{2}\tau)\Big](\beta'-\beta)}{\beta\beta'}\nonumber \\
 & =\frac{\beta'-\beta}{\beta\beta'}\Big(\frac{\pi}{L}\Big)^{-d}\int_{\frac{\pi}{L}}^{\frac{\pi}{a}}d\boldsymbol{q}\Big[1-\exp(-2\nu q^{2}\tau)\Big].\label{eq:average heat}
\end{align}
A cutoff $\pi/a$ of the wavevector is needed to avoid ultra-violet
divergence, i.e., we introduce a smallest spacing $a$ in this elastic
manifold \citep{Huang_book_1987,Parisi_book_1989,Livi_book_2017}.
Since we consider a continuous field, the cutoff spacing is always
much smaller than the system size $a\ll L$. We will see that the
choice of the value of $a$ will influence the average heat (See Fig.
\ref{fig:mean_Q} (b) inset plot).

\begin{figure}
\includegraphics[width=8.5cm]{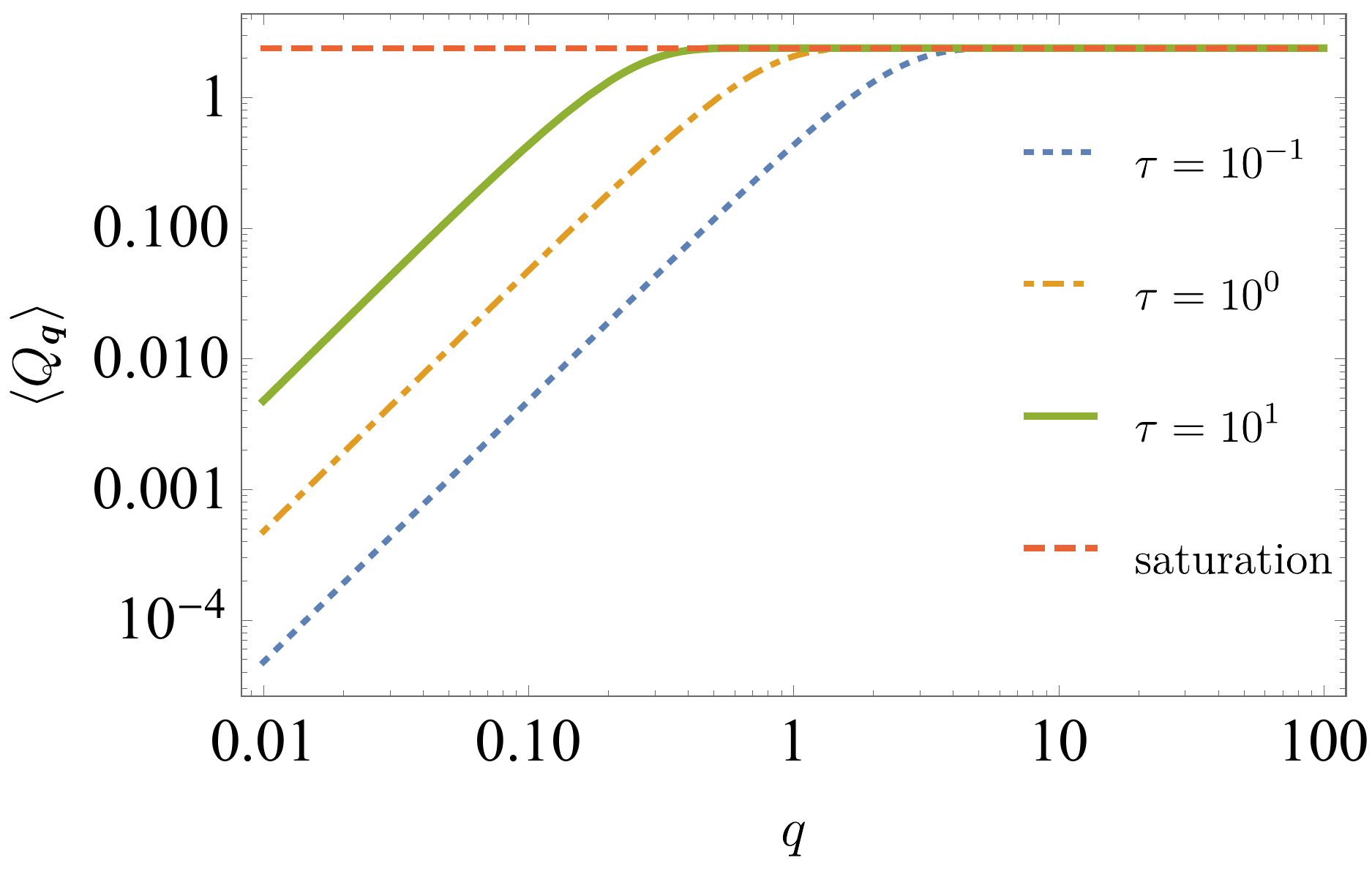}

\caption{{\footnotesize{}Average heat of a mode $\boldsymbol{q}$ for different
time durations. The parameters take values $L=30,\ d=1,\ \nu=1,\ \beta'=4,\ \beta=2$
and the curves correspond to three values of time delay $\tau=10^{1},10^{0},10^{-1}$
from the bottom to the top. The dashed line stands for the saturation
value.\label{fig:log_mean_Q_q}}}
\end{figure}
We rewrite the average heat $\langle Q\rangle$ with a change of the
variable $\boldsymbol{s}=L\boldsymbol{q}$

\begin{align}
\langle Q\rangle & =\frac{(\beta'-\beta)}{\beta\beta'\pi{}^{d}}f\Big(\frac{\nu\tau}{L^{2}}\Big),
\end{align}
where 
\begin{align}
f(r) & =\int_{\pi}^{\frac{L\pi}{a}}d\boldsymbol{s}\Big[1-e^{-2rs^{2}}\Big]\nonumber \\
 & =(\frac{L-a}{a}\pi)^{d}+(\frac{\pi}{8r})^{\frac{d}{2s}}\Big[\mathrm{Erf}(\pi\sqrt{2r})-\mathrm{Erf}(\frac{\pi L\sqrt{2r}}{a})\Big]^{d}.
\end{align}
 $\mathrm{Erf}(r)$ is the error function.

In the following we discuss the asymptotic behavior of the average
heat as a function of time. For one-dimensional case, the average
heat as a function of time is illustrated in Fig. \ref{fig:mean_Q}.
At the initial stage, for $\tau\ll a^{2}/\nu$,

\begin{align}
\langle Q\rangle & \approx\frac{2\pi^{2}}{3a^{2}}\frac{(\beta'-\beta)}{\beta\beta'}\nu\tau\frac{L}{a}.
\end{align}
The average heat initially increases with time linearly. This is Newton's
law of cooling. 

For the intermediate time $a^{2}/\nu\ll\tau\ll L^{2}/\nu$,

\begin{align}
\langle Q\rangle & \approx\frac{(\beta'-\beta)}{\beta\beta'}\frac{L}{a}\Big(1-\frac{a}{\sqrt{8\nu}}\tau^{-1/2}\Big).
\end{align}
It exhibits $\tau^{-1/2}$ scaling with time.

In the long time limit, for $\tau\gg L^{2}/\nu$, 

\begin{align}
\langle Q\rangle & \rightarrow\frac{\beta'-\beta}{\beta\beta'}\frac{L}{a},
\end{align}
the average heat saturates, which is a consequence of the equipartition
theorem. The saturation value of heat is an extensive quantity which
scales linearly with the system size $L$. It will not diverge for
a finite spacing $a$ as a result of finite resolution.

From Eq. (\ref{eq:average heat}) one can see the average heat for
every $\boldsymbol{q}$ mode is 

\begin{equation}
\langle Q_{\boldsymbol{q}}\rangle=\frac{\beta'-\beta}{\beta\beta'}\Big(\frac{\pi}{L}\Big)^{-d}\Big[1-\exp(-2\nu q^{2}\tau)\Big].\label{eq:mean_Q_q}
\end{equation}
 As we can see from this equation and Fig. \ref{fig:log_mean_Q_q},
heat transfer occurs mainly through high-energy modes and occurs in
high-energy modes more quickly than that in lower ones.

For fixed time duration $\tau$, in the small wavevector limit, i.e.,
$2\nu q^{2}\tau\ll1$, it increases with time linearly
\begin{equation}
\langle Q_{\boldsymbol{q}}\rangle=2\nu\tau\frac{\beta'-\beta}{\beta\beta'}\Big(\frac{\pi}{L}\Big)^{-d}q^{2},
\end{equation}
which is the Newton's law of cooling.

On the other hand, if one takes the large wavevector limit, i.e.,
$2\nu q^{2}\tau\gg1$, the average heat reaches the asymptotic value 

\begin{equation}
\langle Q_{\boldsymbol{q}}\rangle=\frac{\beta'-\beta}{\beta\beta'}\Big(\frac{\pi}{L}\Big)^{-d},
\end{equation}
which is the result of the equipartition theorem.

\begin{figure}
\begin{raggedright}
(a)\\
\par\end{raggedright}
\begin{raggedright}
\includegraphics[width=8.5cm]{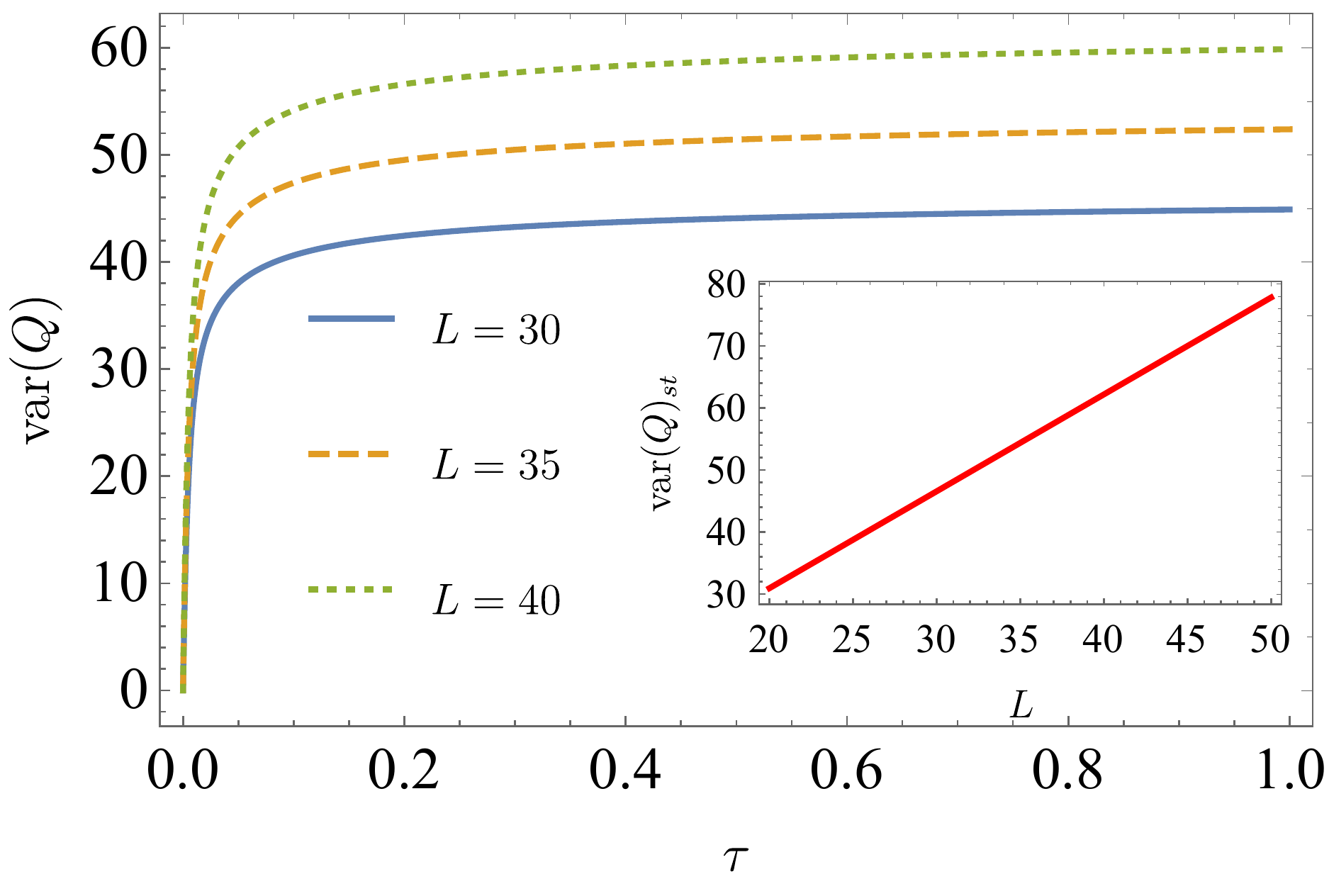}
\par\end{raggedright}
\begin{raggedright}
(b)\\
\par\end{raggedright}
\includegraphics[width=8.5cm]{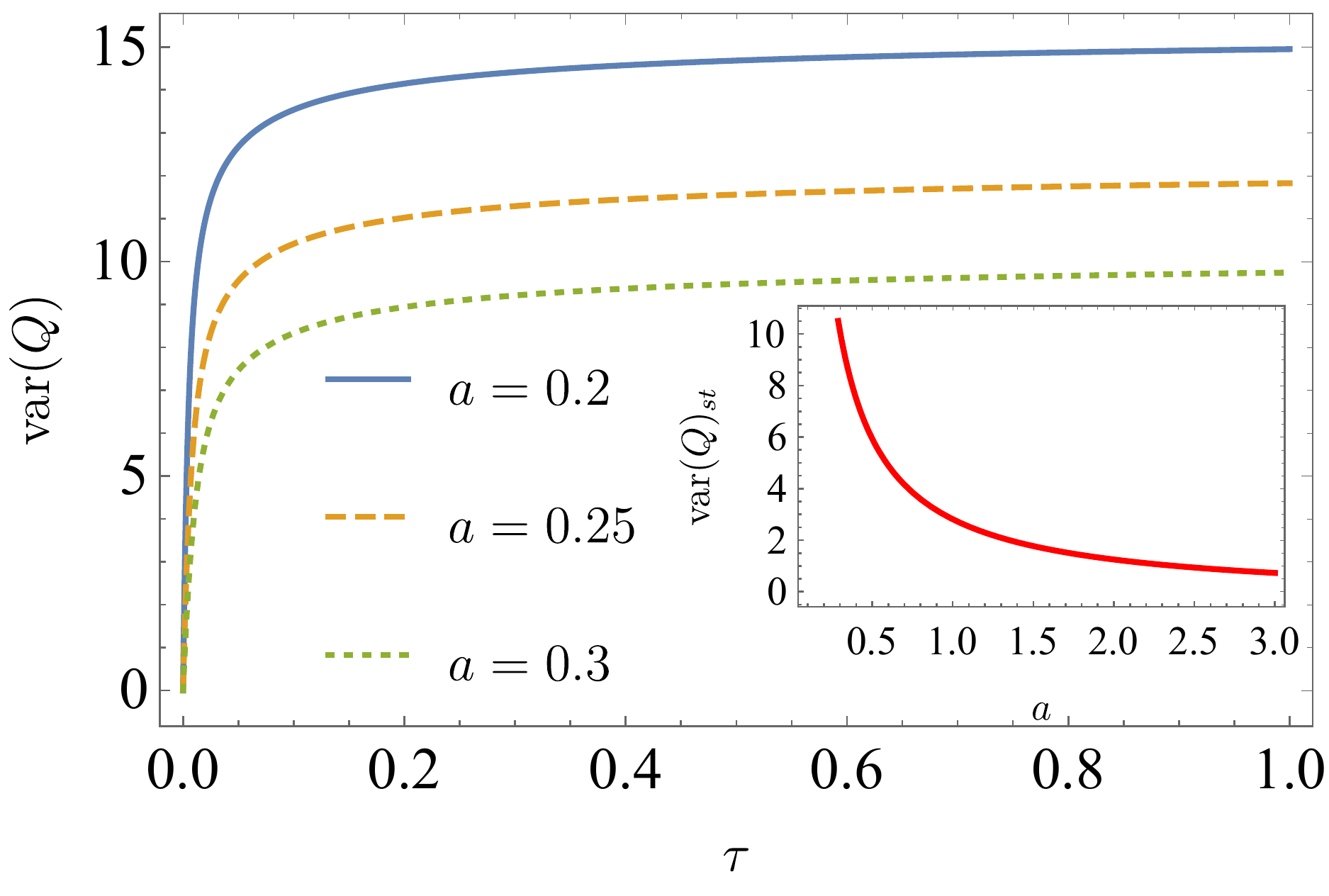}

\caption{{\footnotesize{}Variance of heat as a function of time. Parameters
for both panels: $d=1,\ \nu=1,\ \beta'=4,\ \beta=2.$ (a) $\mathrm{var}(Q)$
as a function of $\tau$ for three system sizes $L=30,35,40$, fixing
$a=0.2.$ Inset: the saturation value of heat variance $\mathrm{var}(Q)_{st}$
as a function of system size $L$. (b) $\mathrm{var}(Q)$ as a function
of $\tau$ for three cutoff spacings $a=0.2,0.25,0.3$, fixing $L=10.$
Inset: the saturation value of heat variance $\mathrm{var}(Q)_{st}$
as a function of cutoff spacing $a$.\label{fig:Variance-of-heat}}}
\end{figure}

From the analytical result of heat statistics Eq. (\ref{eq:GF_of_heat_result})
we can also study the variance of heat. The variance of heat is defined
as $\mathrm{var}(Q)=\langle Q^{2}\rangle-\langle Q\rangle^{2}$, and
can be calculated as 

\begin{align}
\mathrm{var}(Q) & =\frac{1}{i^{2}}\frac{d^{2}\ln\chi_{\tau}(u)}{du^{2}}|_{u=0}\nonumber \\
 & =\Big(\frac{\pi}{L}\Big)^{-d}\frac{1}{\beta^{2}\beta'^{2}}\int_{\frac{\pi}{L}}^{\frac{\pi}{a}}d\boldsymbol{q}e^{-4\nu q^{2}\tau}(-1+e^{2\nu q^{2}\tau})\nonumber \\
 & \quad\bigg[(-1+e^{2\nu q^{2}\tau})\beta^{2}+2\beta\beta'+(-1+e^{2\nu q^{2}\tau})\beta'^{2}\bigg]\nonumber \\
 & =\frac{1}{\beta^{2}\beta'^{2}\pi^{d}}g\Big(\frac{\nu\tau}{L^{2}}\Big)
\end{align}
where

\begin{align*}
g(r) & =\int_{\pi}^{\frac{L\pi}{a}}d\boldsymbol{s}\Big[(\beta^{2}+\beta'^{2})(1-2e^{-2rs^{2}}+e^{-4rs^{2}})\\
 & \quad+2\beta\beta'(-e^{-4rs^{2}}+e^{-2rs^{2}})\Big].
\end{align*}
In the one-dimensional case, for $\tau\ll a^{2}/\nu$, we have

\begin{equation}
\mathrm{var}(Q,\tau)\approx\frac{4\pi^{2}}{3a^{2}\beta\beta'}\nu\tau\frac{L}{a}.
\end{equation}
It grows with time linearly in the very beginning.

For $a^{2}/\nu\ll\tau\ll L^{2}/\nu$, 

\begin{align}
\mathrm{var}(Q,\tau) & \approx\frac{4\pi^{4}\nu^{2}\tau^{2}}{5\beta^{2}\beta'^{2}a^{4}}(\beta^{2}-3\beta\beta'+\beta'^{2})\frac{L}{a}.
\end{align}
It scales as $\tau^{2}$ as time elapses.

Finally, for $\tau\gg L^{2}/\nu$, it reaches the saturation value
in the long time, 

\begin{align}
\mathrm{var}(Q,\tau) & \approx\frac{\beta^{2}+\beta'^{2}}{\beta^{2}\beta'^{2}}\frac{L}{a}.
\end{align}
 As can be seen from Fig. \ref{fig:Variance-of-heat}, the variance
of heat depends on the cutoff spacing $a$ as well. Similar to the
average heat, the saturation value of variance increases linearly
with the system size $L$ and will not diverge for finite spacing
$a$. Higher order cumulants of heat can be analyzed in a similar
way. 

\subsection{Large deviation rate function\label{subsec:Large-deviation-rate}}

We can also study the large deviation rate function of the heat statistics
in the large size limit.

The scaled cumulant generating function (SCGF) $\phi(u,\tau)$ of
heat per volume over time $\tau$, which is defined through

\begin{equation}
\langle\exp[(2L)^{d}u\frac{Q}{(2L)^{d}}]\rangle\asymp_{L\to\infty}e^{(2L)^{d}\phi(u,\tau)}
\end{equation}
or

\begin{align}
\phi(u,\tau) & =\underset{L\to\infty}{\lim}\frac{1}{(2L)^{d}}\ln\langle\exp[(2L)^{d}u\frac{Q}{(2L)^{d}}]\rangle\nonumber \\
 & =\underset{L\to\infty}{\lim}\frac{1}{(2L)^{d}}\ln\chi_{\tau}(-iu),
\end{align}
can be computed by \begin{widetext}

\begin{align*}
\phi(u,\tau) & =\underset{L\to\infty}{\lim}-\frac{1}{(2\pi)^{d}}\int_{\frac{\pi}{L}}^{\frac{\pi}{a}}d\boldsymbol{q}\ln\Big(\frac{-u(\beta'-\beta+u)}{\beta\beta'}\Big[1-\exp(-2\nu q^{2}\tau)\Big]+1\Big).
\end{align*}
\end{widetext}The large deviation rate function for heat per volume
over time $\tau$ is just the Legendre-Fenchel transform of the SCGF
\citep{Touchette_PhysRep_2009}

\begin{align}
I(\frac{Q}{(2L)^{d}},\tau) & =\underset{L\to\infty}{\lim}-\frac{1}{(2L)^{d}}\ln{\cal P}(\frac{Q}{(2L)^{d}},\tau)\nonumber \\
 & =\underset{u\mathbb{\in R}}{\sup}\Big\{ u\frac{Q}{(2L)^{d}}-\phi(u,\tau)\Big\}.\label{eq:LD rate function}
\end{align}

We emphasize that the large deviation rate function of work distribution
in the large size limit has been studied in other models previously
(See e.g., Refs. \citep{Gambassi_PhysRevLett_2012,Hartmann_PhysRevE_2014}).
But as far as we know, the large deviation function of heat in the
large size limit has not been reported previously.

With the large deviation rate function Eq. (\ref{eq:LD rate function}),
we can write down the probability distribution of heat per volume
over time $\tau$ as 

\begin{equation}
{\cal P}(\frac{Q}{(2L)^{d}},\tau)\asymp_{L\to\infty}\exp\Big[-(2L)^{d}I(\frac{Q}{(2L)^{d}},\tau)\Big],
\end{equation}
which demonstrates the dependence of the heat distribution on the
system size. And the fluctuation theorem of heat exchange Eq. (\ref{eq:FT of heat exchange})
can also be formulated in terms of the large deviation rate function.

\section{Conclusion\label{sec:Conclusion}}

Previously, the stochastic thermodynamics of systems with a few degrees
of freedom have been studied extensively both in classical and quantum
realms \citep{Jarzynski_PhysRevLett_1997,Mazonka_Arxiv_1999,Narayan_JPhysA_2003,Speck_PhysRevE_2004,Zon_PhysRevE_2004,Lua_JPhysChemB_2005,Speck_EuroPhysJB_2005,Taniguchi_JStatPhys_2006,Imparato_PhysRevE_2007,Quan_PhysRevE_2008,Engel_PhysRevE_2009,Fogedby_JPhysA_2009,Minh_PhysRevE_2009,Chatterjee_PhysRevE_2010,GomezSolano_PhysRevLett_2011,Nickelsen_EuroPhysJB_2011,Speck_JPhysA_2011,Kwon_PhysRevE_2013,JimenezAquino_PhysRevE_2013,Ryabov_JPhysA_2013,Jarzynski_PhysRevX_2015,Salazar_JPhysA_2016,Zhu_PhysRevE_2016,Funo_PhysRevLett_2018,Funo_PhysRevE_2018,Hoang_PhysRevLett_2018,Pagare_PhysRevE_2019,Fogedby_JStatMech_2020,Chen_Entropy_2021,Gupta_PhysRevE_2021,Chen_PhysRevE_2023,Paraguassu_PhysicaA_2023}.
However, less is known in systems with many degrees of freedom. What
new results the complexity of many degrees of freedom will bring to
stochastic thermodynamics remains largely unexplored.

In this article, we extend previous studies about the stochastic thermodynamics
of systems with a few degrees of freedom to a continuous field. We
compute the heat statistics in the relaxation process of an exactly
solvable model \textemdash{} an elastic manifold whose underlying
dynamics can be described by the Edwards-Wilkinson equation. By employing
Feynman-Kac approach, we calculate analytically the characteristic
function of heat for any relaxation time. The analytical results of
heat statistics have pedagogical value and may bring important insights
to the understanding of thermodynamics in extreme nonequilibrium processes.
For example, the cumulants of heat in such a system with many degrees
of freedom require a spatial cutoff to avoid the ultra-violet divergence,
which is a consequence of finite resolution. We also analyze the scaling
behavior of the cumulants with time and the system size. In addition,
the large deviation rate function of heat in the large size limit
is analyzed.

This work can be regarded as an early step in the stochastic thermodynamics
of continuous fields. More interesting problems remain to be explored
such as the definitions for the thermodynamic quantities in every
space-time point, the extension to nonlinear models, the work statistics
in the presence of external driving and so on. Studies about these
issues will be given in our future work.
\begin{acknowledgments}
This work is supported by the National Natural Science Foundation
of China (NSFC) under Grants No. 12147157, No. 11775001, and No. 11825501.
\end{acknowledgments}

\appendix

\section{Derivation of Eq. (\ref{eq:=00005Ceta_t})\label{sec:Derivation}}

Similar to the probability density distribution, the modified function
$\eta(h,t)$ can be written as the product of the imaginary part and
the real part over all modes in the Fourier space

\begin{equation}
{\cal \eta}(\{h_{\boldsymbol{q}}\},t)=\underset{q_{i}\geq\pi/L}{\prod}{\cal \eta}_{\boldsymbol{q}}(h_{\boldsymbol{q}}^{R},t){\cal \eta}_{\boldsymbol{q}}(h_{\boldsymbol{q}}^{I},t).
\end{equation}
The probability-density-like function $\eta(h,t)$ follows the same
time evolution as ${\cal P}(h,t)$ in Eq. (\ref{eq:Fokker-Planck Fourier}) 

\begin{align}
\frac{\partial{\cal \eta}_{\boldsymbol{q}}(h_{\boldsymbol{q}}^{R,I},t)}{\partial t} & =\frac{(2\pi)^{d}}{2\beta}\frac{\partial^{2}{\cal \eta}_{\boldsymbol{q}}}{\partial(h_{\boldsymbol{q}}^{R,I})^{2}}+\nu q^{2}{\cal \eta}_{\boldsymbol{q}}+\nu q^{2}h_{\boldsymbol{q}}^{R,I}\frac{\partial{\cal \eta}_{\boldsymbol{q}}}{\partial h_{\boldsymbol{q}}^{R,I}}.
\end{align}
with the initial condition 

\begin{equation}
\eta(h,0)=e^{-iuH_{S}(0)}{\cal P}(h,0).\label{eq:initial_eta}
\end{equation}
Due to the quadratic nature of the EW equation, we assume the time-dependent
solution $\eta(h,t)$ takes a Gaussian form at any time

\textbf{
\begin{align}
{\cal \eta}_{\boldsymbol{q}}(h_{\boldsymbol{q}}^{R,I},t) & =\sqrt{\frac{\beta'\nu q^{2}}{\pi(2\pi)^{2d}}}\exp\Big[-A(t)(h_{\boldsymbol{q}}^{R,I})^{2}+B(t)\Big].\label{eq:=00005Ceta_AtBt}
\end{align}
}The coefficients are governed by the following ordinary differential
equations

\begin{equation}
\dot{A}(t)=-\frac{2(2\pi)^{d}}{\beta}A^{2}(t)+2A(t)\nu q^{2},
\end{equation}

\begin{equation}
\dot{B}(t)=-\frac{(2\pi)^{d}}{\beta}A(t)+\nu q^{2}.
\end{equation}
The initial condition Eq. (\ref{eq:initial_eta}) gives way to the
initial values of the coefficients

\begin{equation}
A(0)=(\beta'+iu)\nu\frac{1}{(2\pi)^{d}}q^{2},
\end{equation}

\begin{equation}
B(0)=0.
\end{equation}
By solving the above equations we obtain

\begin{align}
A(t) & =\frac{1}{(2\pi)^{d}}\frac{e^{2\nu q^{2}t}\beta(u-i\beta')\nu q^{2}}{(e^{2\nu q^{2}t}-1)u-i[\beta+(e^{2\nu q^{2}t}-1)\beta']},\label{eq:A(t)}\\
B(t) & =\nu q^{2}t+\frac{1}{2}\ln\left[\frac{i\beta}{u-i\beta'+i\beta+(i\beta'-u)e^{2\nu q^{2}t}}\right].\label{eq:B(t)}
\end{align}
Substituting Eqs. (\ref{eq:A(t)}) and (\ref{eq:B(t)}) into Eq. (\ref{eq:=00005Ceta_AtBt}),
we arrive at \begin{widetext}

\begin{align}
{\cal \eta}(\{h_{\boldsymbol{q}}\},t) & =\underset{q_{i}\geq\pi/L}{\prod}{\cal \eta}_{\boldsymbol{q}}(h_{\boldsymbol{q}}^{R},t){\cal \eta}_{\boldsymbol{q}}(h_{\boldsymbol{q}}^{I},t)\nonumber \\
 & =\underset{\boldsymbol{q}(q_{i}\geq\pi/L)}{\prod}\frac{\beta'\nu q^{2}}{\pi(2\pi)^{d}}\frac{i\beta\exp(2\nu q^{2}\tau)}{u-i\beta'+i\beta+(i\beta'-u)\exp(2\nu q^{2}t)}\nonumber \\
 & \hfill\exp\bigg\{-\frac{1}{(2\pi)^{d}}\frac{\exp(2\nu q^{2}t)\beta(u-i\beta')\nu q^{2}}{\Big[-1+\exp(2\nu q^{2}t)\Big]u-i\Big[\beta-\beta'+\beta'\exp(2\nu q^{2}t)\Big]}\Big[(h_{\boldsymbol{q}}^{R})^{2}+(h_{\boldsymbol{q}}^{I})^{2}\Big]\bigg\}.\label{eq:=00005Ceta_t}
\end{align}
\end{widetext}

Substituting it into Eq. (\ref{eq:GF_heat_general}), we obtain the
characteristic function of heat Eq. (\ref{eq:GF_of_heat_result})
of the EW elastic manifold in the relaxation process.

\bibliographystyle{apsrev4-1}
\bibliography{D:/dadadadida/NEQ/EW/EW_ref}

\end{document}